\documentclass[twocolumn,aps,prd,unsortedaddress,superscriptaddress,a4paper,nofootinbib,balancelastpage,preprintnumbers,showpacs,floatfix]{revtex4-1}
\usepackage{latexsym,amsbsy,amsmath}
  \usepackage[pdftex]{graphicx}
\usepackage{longtable}
\usepackage{dcolumn}
 \DeclareGraphicsExtensions{.eps,.jpg,.png}
\renewcommand{\vec}[1]{\boldsymbol{#1}}
\begin{document}
%\title{Dibaryion Vs. baryonium in the light sector. A multiquark perspective}
%\title{Multiquark contributions to dibaryons and baryonium}
\title{Adiabaticity and color mixing in tetraquark spectroscopy}
\author{J.~Vijande}
\email{javier.vijande@uv.es}
\affiliation{Departamento de F\'{\i}sica At\'{o}mica, Molecular y Nuclear, Universidad de Valencia (UV)
and IFIC (UV-CSIC), Valencia, Spain.}
\author{A. Valcarce}
\email{valcarce@usal.es}
\affiliation{Departamento de F{\'\i}sica Fundamental,
Universidad de Salamanca, 37008 Salamanca, Spain}
\author{J.-M.~Richard}
\email{j-m.richard@ipnl.in2p3.fr}
\affiliation{Universit\'e de Lyon, Institut de Physique Nucl\'eaire de Lyon,
IN2P3-CNRS--UCBL,\\
4 rue Enrico Fermi, 69622  Villeurbanne, France}
\date{\today}
\begin{abstract}
 We revisit the role of color mixing in the quark-model calculation of tetraquark states, and compare simple pairwise potentials to more elaborate string models with three- and four-body forces. We attempt to disentangle the improved dynamics of confinement from the approximations made in the treatment of the internal color degrees of freedom.
\end{abstract}
\maketitle
\section{Introduction}
There is a persisting interest in the quark dynamics applied to multiquark spectroscopy. The question is whether there exist compact hadron states beyond ordinary mesons (quark--antiquark) and baryons (three quarks). 

In simple constituent models, several mechanisms have  been proposed for binding multiquarks: chiral dynamics for light quarks, chromomagnetism, clustering of heavy quarks in a chromoelectric potential, etc. In this article, we concentrate on this latter effect, i.e., multiquark binding in a spin- and flavor-independent potential,   and discuss the role of the  internal color degrees of freedom. The validity of the existing  models is beyond the scope of this note. In particular, we shall not discuss the transition from color as a local gauge invariance to color as a global property of the wave-function in constituent models. Still, even in its simplified version, color is a delicate ingredient of multiquark dynamics.

The first studies on multiquarks  within constituent models were based on simple color-additive potentials, extrapolated from meson and baryon spectroscopy. Already for baryons, one hardly justifies the choice of a pairwise interaction, half as strong as the quark--antiquark potential, but a more elaborate modeling, based on a connected $Y$-shape flux tube linking the three quarks, does not change the results significantly.

We shall follow in this paper this picture of a minimal string linking the quarks. There are alternative non-trivial pictures of confinement, in particular the ones based on diquarks, which have been extended from the baryon sector to the multiquarks. See, for instance, \cite{Maiani:2004vq,Ebert:2005nc,Dubnicka:2010kz}.

The $Y$-shape potential of baryons has been extended to tetraquarks and higher multiquark configurations \cite{Carlson:1991zt}. There are multi-$Y$ connected diagrams in which the string interaction links all quarks and antiquarks as a Steiner tree whose cumulated length is minimized. But the dynamics is dominated by the so-called ``flip-flop'' diagrams, with disconnected flux tubes for each of the quark--antiquark, three-quark or three-antiquark sub-clusters: the attraction comes from the minimum taken over all possible permutations of the quarks and of the antiquarks. 

This flip--flop interaction contains 3-body, 4-body, and higher-order terms, and is thus more delicate to handle in variational calculations. Moreover, when two quarks are exchanged, the color wave function is modified. In the latest studies \cite{Vijande:2007ix,Richard:2009rp,Vijande:2011im}, this effect is not treated rigorously. Instead, a type of adiabatic approximation is used: for any set of coordinates for the quarks, the potential is taken as the minimum of all permutations of the quarks and antiquarks, irrespective of the color wave function, and  this minimum, as a function of the coordinates, is interpreted as an effective potential leading to a few-body spectral problem in which color has disappeared. Interestingly, this strategy leads to stable multiquarks for a large variety of constituent masses. This is at variance with the color-additive model, which binds only tetraquarks for large values of the quark-to-antiquark mass ratio.

The question is thus whether the multiquark binding obtained in  string models survives a non-adiabatic treatment of the internal color degrees of freedom. The problem is analyzed in the present paper by reformulating the string-based interaction as an operator in color space.

The outline is the following. In Sec.~\ref{se:color-wf}, we set the notation for the color components of tetraquarks. In Sec.~\ref{se:models}, we present various models of tetraquark confinement with coupled channels in color space or an adiabatic approximation in which color disappears from the final wave-equation. The results are presented in Sec.~\ref{se:resu}, and the conclusions in Sec.\ref{se:conc}.
\section{Color states}\label{se:color-wf}
The internal color structure of tetraquark states is described in several papers, see, e.g. \cite{Chan:1977st,Chan:1978nk,Vijande:2009ac}. We shall borrow the notation of \cite{Chan:1977st}, in particular the names ``true baryonium''  ($T$) and ``mock baryonium'' ($M$), though the physics context is rather different in the present heavy-quark spectroscopy as compared to the color chemistry of the late 70s.

The wave function for $(1,2,3,4)=(qq\bar q\bar q)$ is written as 
\begin{equation}\label{eq:col-tetra}
\Psi=\psi_T\,|T\rangle+\psi_M\,|M\rangle
\end{equation}
where $|T\rangle$ denotes a color state with the two quarks in a color $\bar 3$ state, and the antiquarks in a color 3, while $|M\rangle$ corresponds to a color sextet in the quark sector and antisextet in the antiquark one. There is also the possibility of building the global color state out of quark--antiquark clusters either in a color singlet or octet. More precisely, we introduce
\begin{equation}\label{eq:color-st}
\begin{aligned}
 |T\rangle&=|(12)_{\bar 3}\,(34)_3\rangle~,\quad& |M\rangle&=|(12)_{6}\,(34)_{\bar 6}\rangle~,\\
|1\phantom{'}\rangle&=|(13)_{1}\,(24)_1\rangle~,\quad& |8\phantom{'}\rangle&=|(13)_{8}\,(24)_{8}\rangle~,\\
|1'\rangle&=|(14)_{1}\,(23)_1\rangle~,\quad& |8'\rangle&=|(14)_{8}\,(23)_{8}\rangle~.\\
\end{aligned}
\end{equation}

The relations between the different sets can be deduced from
\begin{equation}\label{eq:color-rel}
\begin{aligned}
 |1\phantom{'}\rangle&=\phantom{+}\sqrt{\frac13}\,|T\rangle+\sqrt{\frac23}\,|M\rangle~,\\
|8\phantom{'}\rangle&=-\sqrt{\frac23}\,|T\rangle+\sqrt{\frac13}\,|M\rangle~,\\
 |1'\rangle&=-\sqrt{\frac13}\,|T\rangle+\sqrt{\frac23}\,|M\rangle~,\\
|8'\rangle&=\phantom{+}\sqrt{\frac23}\,|T\rangle+\sqrt{\frac13}\,|M\rangle~.\\
\end{aligned}
\end{equation}

Accordingly, the matrix elements of the potential in any basis are related to the ones in another basis, for instance,
\begin{equation}\label{eq:pot-rel}
V_{11}=\langle 1|V|1\rangle =\frac13\,V_{TT}+\frac23\,V_{MM}+\frac{2\,\sqrt{2}}{3}\,V_{TM}~,
\end{equation}
and many similar relations. 

As stressed by Lipkin \cite{Lipkin:1986dw}, in the limit of a tetraquark with two units of heavy flavor, $(QQ\bar q\bar q)$ with a large quark-to antiquark mass ratio $M/m$, the ground state is an almost pure $|T\rangle$ state, with the two flavored quarks in an antitriplet state, as in ordinary $(QQq)$ baryons, and the two antiquarks neutralizing that color, as in $(\bar Q\bar q\bar q)$ antibaryons. In other words, the tetraquark state in the large $M/m$ limit just uses well probed color structures, such as the $3\otimes 3\to \bar 3$ coupling of two quarks in baryons.

On the other hand, for smaller  values of the mass ratio $M/m$,   the simple models give at best a very shallow binding. Then the mixing of $|T\rangle$ and $|M\rangle$ is crucial to establish the stability. See, e.g., Brink and Stancu \cite{Brink:1994ic}.
\section{Models of tetraquark confinement}\label{se:models}
In early days of multiquark calculations, the potential was assumed to be pairwise, with the color dependence associated with the exchange of a color octet, namely
\begin{equation}\label{eq:add}
 V=-\frac{3}{16}\,\sum_{i<j} \tilde\lambda_i.\tilde\lambda_j\,v(r_{ij})~.
\end{equation}
Here, $\tilde\lambda_i$ is the color operator for the $i^\text{th}$ quark, and is suitably modified for an antiquark belonging to the $\bar 3$ representation of SU(3). The normalization is such that $v(r_{ij})$ is the central part of the quarkonium potential. 

This model, however crude, gives at least the possibility of studying the role of the internal color degrees of freedom inside a multiquark state.

\begin{figure}[!!!ht]
 \centering
 \raisebox{.5cm}{\includegraphics[scale=.9]{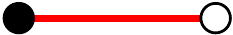}}\hfil
\includegraphics[scale=.9]{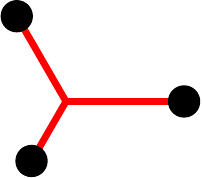}\\[10pt]
\includegraphics[scale=.9]{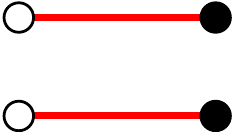}
\qquad \includegraphics[scale=.9]{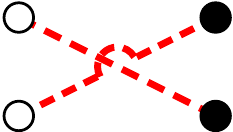}
\qquad \includegraphics[scale=.9]{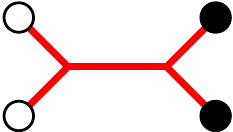}
 \caption{Top: schematic picture of the quark--antiquark and three-quark confinement. Bottom: three contributions to the tetraquark potential; the simple string model takes the minimum of these three contributions}
 \label{fig:mes-bar-tetra}
\end{figure}

The additive model being subject to heavy criticism, an alternative was sought, inspired by the strong-coupling regime of QCD. It is referred to as the \emph{flip--flop} model. For mesons, the potential is $V=\sigma\,r$, where $r$ is the quark--antiquark distance, and $\sigma$ the string tension, which can be set to $\sigma=1$ by rescaling, without loss of generality. For baryons, this is the $Y$-shape interaction $V_Y=\min_a(r_{1a}+r_{2a}+r_{3a})$. For a tetraquark, the potential is the minimum of the flip--flop term and a connected string, namely \cite{Carlson:1991zt,Vijande:2007ix}
\begin{equation}\label{eq:pot-tetra}
\begin{aligned}
V_4&=\min(V_\text{ff},V_\text{s})~,\\
V_\text{ff}&=\min(r_{13}+r_{24},r_{14}+r_{23})~,\\
V_\text{s}&=\min_{a,b}(r_{1a}+r_{2a}+r_{ab}+r_{3b}+r_{4b})~,
\end{aligned}
\end{equation}
as pictured in Figs.\ \ref{fig:mes-bar-tetra}.

For each set of quark coordinates $\vec r_i$, the minimization in \eqref{eq:pot-tetra} implies a rotation in color space. This means that  $V_4$ is an adiabatic approximation, which tends to overbind the system. More importantly, the minimization, at least as it was carried out in \cite{Vijande:2007ix}, does not account for any antisymmetrization. It just holds for distinguishable quarks or antiquarks. 

As the model \eqref{eq:pot-tetra} gives an interesting spectrum of stable tetraquarks \cite{Vijande:2007ix}, and, if extended to higher configurations, a spectrum of bound pentaquarks and hexaquarks \cite{Richard:2009rp,Vijande:2011im}, it is crucial to estimate the amount of overbinding due to the adiabatic approximation, and the changes occurring when a proper antisymmetrization is implemented.

We aim at constructing an operator in color space that tends to \eqref{eq:pot-tetra} in the adiabatic limit, at least when one of the three terms is clearly the minimum. For instance, if the (1,3)  pair is clustered and lies far from the (2,4) pair  which is also clustered, the potential is more easily described in the $\{|1\rangle,\,|8\rangle\}$ basis. 
However, working  solely in this latter basis would require the choice of  a value for the string tension between color-octet objects (there are studies within QCD, see, e.g.,~\cite{Bali:2000un}) and  an ansatz for the transition potential $V_{18}$. Instead, we shall combine the pieces of information coming from the singlet--singlet and triplet--antitriplet states, to deduce the full $2\times2$ matrix of the potential in the $\{|T\rangle,\,|M\rangle\}$ basis, in which the four-body problem will be solved. 

More precisely, we will consider four different models:
\begin{itemize}
 \item 
 Model A is the adiabatic limit given by  \eqref{eq:pot-tetra}, already used in \cite{Vijande:2007ix}. However, the Steiner-tree with two junctions, that plays a marginal role, is neglected. Hence, this is the pure flip--flop model.
% \item
% Model 2 uses simultaneously $V_{11}$, ,$V_{1'1'}$ and $V_{TT}$ to reconstruct the missing $V_{TM}$ and $V_{MM}$, with the result
% %
% \begin{equation}\label{eq:model2}
%  \begin{aligned}
%   V_{TM}&=\frac{3}{4\sqrt{2}}\left(V_{11}-V_{1'1'}\right)\\
%   V_{MM}&=\frac{1}{4}\left(3V_{11}+3V_{1'1'}-2V_{33} \right)~,
%  \end{aligned}
% \end{equation}
% %
% where  $V_{11}=r_{13}+r_{24}$, $V_{1'1'}=r_{14}+r_{23}$ and, for the sake of convenience, the Steiner tree in $V_{TT}$ is approximated by a rectangular diagram, namely, 
% %
% \begin{equation}\label{eq:model2a}
% V_{TT}=r_{12}+r_{34}+\frac{1}{2}|(\vec r_1+\vec r_2)-(\vec r_3+\vec r_4)|~.
% \end{equation}
% %
% This is obviously more attractive than our model~1. For instance, in a configuration where $V_{11}$ is the lowest term of the flip-flop, model 1 takes this $V_{11}$ as the potential, whereas  model 2 provides it with more attraction by means of a coupling to another channel.
\item
Model B is a smooth version of the adiabatic approximation. We use $g(x)=1/(1+x^n)$ and its complement $g'(x)=1-g(x)$,  with some large but finite exponent $n=5$ to soften the transition between different limiting regimes of the string model. In practice, we replace the above $V_{TT}$  by 
\begin{multline}\label{eq:model3}
 \hat V_{TT}= g_3\,g'_3\, V_{TT}+(1-g_3\,g'_3)\\
{}\times
\Biggl[\frac{3\,V_{1'1'}+V_{11}}{4}\,g_1
{}+\frac{3\,V_{11}+V_{1'1'}}{4}\,g'_1\Biggr]~,
\end{multline}
where
 \begin{equation}\label{eq:model3p}
 V_{11}=r_{13}+r_{24}~,\quad V_{1'1'}=r_{14}+r_{23}~, 
 \end{equation}
and
\begin{equation}\label{eq:model3b}
\begin{aligned}
 g_3&=g\left(\frac{V_{33}}{V_{11}}\right)~,\quad
g'_3=g\left(\frac{V_{33}}{V_{1'1'}}\right)~,\\
g_1&=1-g'_1=g\left(\frac{V_{11}}{V_{1'1'}}\right)~.
\end{aligned}
\end{equation}

% \begin{multline}\label{eq:model3}
%  \hat V_{TT}= g\left(\frac{V_{33}}{V_{11}}\right)g\left(\frac{V_{33}}{V_{1'1'}}\right)\, V_{TT}\\
% {}+\left[1-g\left(\frac{V_{33}}{V_{11}}\right)g\left(\frac{V_{33}}{V_{1'1'}}\right)\right]\\
% {}\times
% \Biggl[\frac{3\,V_{1'1'}+V_{11}}{4}\,g\left(\frac{V_{11}}{V_{1'1'}}\right)\\
% {}+\frac{3\,V_{11}+V_{1'1'}}{4}\,g\left(\frac{V_{1'1'}}{V_{11}}\right)\Biggr]~.
% \end{multline}
%
Then,  the potential is known in any basis from $V_{11}$, $V_{1'1'}$ and $\hat V_{TT}$. For instance, in  the $\{|1\rangle,\,|8\rangle\}$ basis, one uses $V_{11}$ and 
\begin{equation}
 \label{eq:model3c}
\begin{aligned}
  V_{18}&=\frac{1}{4\sqrt{2}}\{V_{11}+3V_{1'1'}-4\hat V_{TT}\}\\
  V_{88}&=\frac{1}{4}\{-V_{11}+3V_{1'1'}+2\hat V_{TT}\}~,
\end{aligned}
\end{equation}
and it is readily checked that if $g_1\to 1$, the potential becomes diagonal. The four-body problem is more conveniently solved in the $\{|T\rangle,\,|M\rangle\}$ basis, and besides $\hat V_{TT}$, the relevant matrix elements are
\begin{equation}
 \label{eq:model3a}
\begin{aligned}
  V_{TM}&=\frac{3}{4\sqrt{2}}\{V_{11}-V_{1'1'}\}\\
  V_{MM}&=\frac{1}{4}\{3V_{11}+3V_{1'1'}-2\hat V_{TT}\}~,
\end{aligned}
\end{equation}
\item 
Model C is the color-additive model of \eqref{eq:add}, normalized to a unit string tension for quarkonium. 
\item Model D is the crude adiabatic limit of the previous one. This means that for any given set of positions, the $2\times2$ matrix consisting of $V_{TT}$, $V_{TM}$ and $V_{MM}$ of model 4 is diagonalized, and the lowest eigenvalue is taken as the effective four-body potential, irrespective of any symmetrization or antisymmetrization. Of course, model D is more attractive than model~C.
\item Several other variants have been envisaged, but abandoned as leading to collapses or inconsistencies. 
For instance, it is tempting to use relations similar to \eqref{eq:model3a} to express $V_{TM}$ and $V_{MM}$ from $V_{11}$, $V_{1'1'}$ and $V_{TT}$, as given by the string model. But then the Hamiltonian is not bounded below.
\end{itemize}
\section{Results}\label{se:resu}
The results are shown in Table~\ref{tab:resu}. The aim is not to provide a benchmark of four-body calculations. What really matters, is how the ground-state energy evolves when going from a model to another one. 
The variational estimate has been carried out using just a few Gaussians, and in the case of models B and C, imposing the proper symmetry.

For this purpose, we make use of a wave function with the relevant 
symmetry for each color component. As the color vector $|T\rangle$ ($|M\rangle$) is antisymmetric (symmetric) under both the 
exchange of the identical quarks and the identical antiquarks, it has to be combined with a radial wave function with the proper symmetries. 
The way of constructing these wave functions has been explicitly
detailed in Ref.~\cite{Vijande:2009ac}. We will just draft here its main characteristics. The radial wave function is taken as a 
linear combination of generalized Gaussians depending on six variational parameters $a_{ij}$ of the form,
\begin{equation}
\label{fbp:eq:gauss}
\Psi(\vec{x_1},\vec{x_2},\vec{x_3})=\sum_{k=1}^4\alpha_k\exp\left[-\sum_{i\geq j=1}^3 a_{ij}s_{ij}^k\,\vec{x}_i\cdot\vec{x}_j\right],
\end{equation}
where $s^k$ are six-component $(ij)$ vectors made of arrangements of positive and negative signs. Once combined with the proper election of the signs $\alpha_k$, they give rise to radial wave functions with the following symmetries in the radial space under the exchange 
of quarks and antiquarks: $SS,SA,AS,$ or $AA$, where $S$ stands for symmetric and $A$ for antisymmetric.
 \begin{table}[!htbc]
\caption{\label{tab:resu} Results for the various models A, B, C and D, as a function of the quark-to-antiquark mass ratio $M/m$}
\begin{ruledtabular}
\begin{tabular}{c|cccc|c}
$M/m$ & A & B & C & D & Threshold\\
\hline
1. & 4.644 &  4.803 & 4.702 & 4.596 & 4.676\\
2. & 4.211 &  4.306 & 4.275 & 4.160 & 4.248\\
3. & 4.037 &  4.131 & 4.112 & 3.984 & 4.086\\
4. & 3.941 &  4.041 & 4.010 & 3.891 & 3.998\\
5. & 3.880 &  3.985 & 3.954 & 3.828 & 3.942\\
10.& 3.742 &  3.860 & 3.834 & 3.685 & 3.831
\end{tabular}
\end{ruledtabular}
\end{table}

The results from models A and D are very similar for the variational energies. This means that the extra attraction noticed in \cite{Vijande:2007ix}
is mainly due to the color mixing in the adiabatic approximation. Restoring the interaction as an operator in color space is less favorable for multiquark binding, and it is readily seen that models B and C led to comparable predictions.
The color structure of the wave-functions is also rather similar in models B and C, with mostly a singlet--singlet configuration.

However, if one looks at the details, it can be realized that the flip--flop and and the additive models differ. In particular, the average $QQ$ separation, $\langle r_{12}\rangle$ is smaller in the additive model than in the flip--flop model. 
If the mass ratio $M/m\to\infty$, this effect becomes more pronounced, with $\langle r_{12}\rangle\to 0$ in the former case and 
$\langle r_{12}\rangle$ evolving to a finite value (for $m$ fixed) in the latter one.
This influences the contribution of the $W$-exchange contribution to the weak decay of $(bc\bar q\bar q)$, which also plays a role in the decay of $(bcu)$ baryons \cite{Korner:1994nh}.

\begin{figure}
 \includegraphics[scale=.99]{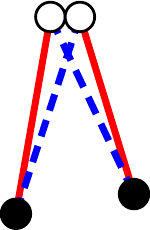}\hspace*{1.5cm}
 \includegraphics[scale=.99]{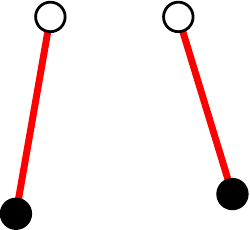}
 % FigsAdia-fig6.pdf: 43x66 pixel, 72dpi, 1.52x2.33 cm, bb=0 0 43 66
 \caption{Flip-flop potential for $r_{12}\to 0$ and cumulated potential of two mesons}
 \label{fig:BO}
\end{figure}

This large $M/m$ limit is rather interesting. A detailed numerical study would involve the computation of an effective $QQ$ interaction, $V_\text{eff}(r_{12})$, sum of the direct $QQ$ interaction and of the $\bar q\bar q$ binding energy around them, very similar to the Heitler--London potential of two protons in the adiabatic treatment of the hydrogen molecule. Already, for baryons with two heavy quarks, $(QQq)$, it was stressed that the Born--Oppenheimer approximation works very well to reproduce the properties of the lowest states \cite{Fleck:1989mb}.  
For $(QQ\bar q\bar q)$, it is striking that the effective potentials $V_\text{eff}(r_{12})$ have qualitative differences in the additive model and in the flip--flop one.  
In the additive model, the minimum of $V_\text{eff}(r_{12})$ is reached at $r_{12}=0.$ The $\bar q\bar q$ contribution is stationary there. So, up to a constant, the potential is dominated by the direct $QQ$ term, which is $r_{12}/2$ here. Thus  the average separation decreases as $M^{-1/3}$, according to the well-known scaling laws in a linear potential \cite{Quigg:1979vr}. In the flip--flop model,  $V_\text{eff}(r_{12})$ is minimum at some finite distance.
This can be seen directly. This is also compulsory, if one wishes to understand why $(QQ\bar q\bar q)$ is stable in the large $M/m$ limit, as seen by the following \textsl{reductio ad absurdum}: suppose that in the Born--Oppenheimer limit of the flip--flop model, $\langle r_{12}\rangle\to 0$, then the flip--flop energy of $(QQ\bar q\bar q)$ and the energy of its threshold, $(Q\bar q)+(Q\bar q)$, which are pictured in Fig.~\ref{fig:BO} and given by
\begin{equation}\label{eq:largeMmpot}
\begin{aligned}
 V_\text{eff}(r_{12})&=\min(r_{13}+r_{24},r_{14}+r_{23})~,\\
V_\text{th}&=r_{13}+r_{24}~,
\end{aligned}
\end{equation}
would coincide if $r_{12}=0$ and  the inequality among energies $(QQ\bar q\bar q)<(Q\bar q)+(Q\bar q)$ would be impossible.

\begin{figure}[b]
 \includegraphics[scale=.99]{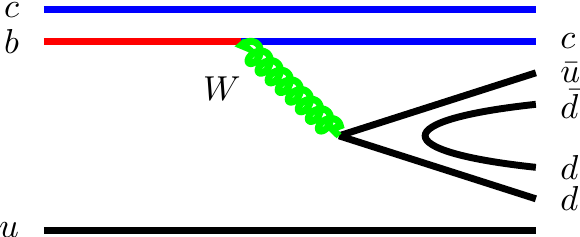}\\[15pt]
 \includegraphics[scale=.99]{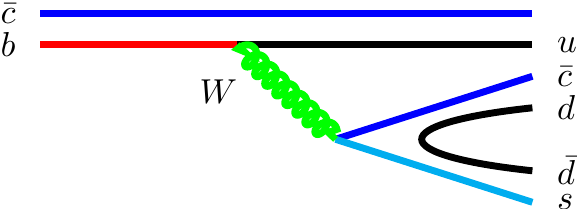}\\[15pt]
\includegraphics[scale=.99]{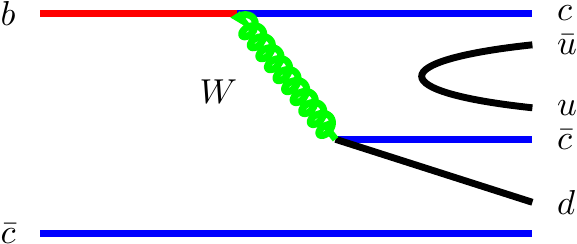}\\
 % FigsAdia-fig6.pdf: 43x66 pixel, 72dpi, 1.52x2.33 cm, bb=0 0 43 66
 \caption{Some weak decays of the $(bcu)$ baryon and $(b\bar c)$ meson leading to a tetraquark in the final state}
 \label{fig:weak}
\end{figure}
\section{Conclusions}\label{se:conc}
Our results illustrate the role of antisymmetrization in preventing from a proliferation of multiquarks. In particular, the main difference between the naive color-additive model and the crude flip--flop model comes from the treatment of color.
It remains that in such a flavor-independent confining, the binding of $(QQ\bar q\bar q)$ is obtained for heavy enough quarks.

More ambitious tools are obviously needed to handle the four-quark dynamics. In particular, in the double-charm sector, the binding effect of the confining interaction is probably not sufficient. In  addition, the spin-part of the wave function can be made favorable: as stressed, e.g., in \cite{Hyodo:2012pm} and references therein, there is a light--light interaction in $(cc\bar u\bar d)$ which is absent in the two mesons constituting the threshold. 

The production of such tetraquarks can be accessible in B-factories \cite{Hyodo:2012pm} and at the proton--proton colliders. Note also that $(cc\bar q\bar q)$ can be  a decay product of hadrons containing charm and beauty. From a $(bcu)$ baryon, for instance, the Cabibbo allowed 
$b\to c+W^-\to c+ d+\bar u$, combined to a $d\bar d$ pair creation, leads to $(bcu)\to (cc\bar u\bar d)+ (ddu)$. See Fig.~\ref{fig:weak}.
From $(b\bar c)$, one could first envisage the Cabbibo suppressed $b\to u+W^-\to u+\bar c+s$, and after the creation of a light quark--antiquark pair, this would monitor a decay $(b\bar c)\to (\bar c\bar cud)+(s \bar d)$. Of course, the CKM suppression factor is rather effective here. Perhaps more promising is the chain $b\to c+W^-\to c+\bar c+d$ giving altogether
$(c\bar c u \bar u d \bar c)$ after a $u\bar u$ pair creation. This could lead to $\bar D^-+X$, where $\bar D$ is an anticharmed meson and $X$ one of the new hidden charm resonances  reviewed, e.g., in \cite{Nielsen:2009uh,Valcarce:2012qwa}. Another combination is
$(\bar c\bar c u d)+ (c\bar u)$, with, however, a different topology of the quark diagram and thus different color and OZI suppression factors, as discussed by Lipkin in a different context \cite{Lipkin:1998ew}. Anyhow, any heavy-quark factory should lead to the discovery of heavy tetraquarks with suitable triggers.

\acknowledgments
We are very grateful to several colleagues for very useful discussions, in particular Makoto~Oka, Paolo Gambino and Qiang Zhao.
This work has been partially funded by the Spanish Ministerio de
Educaci\'on y Ciencia and EU FEDER under Contracts No. FPA2010-21750
and AIC-B-2011-0661,
by the Spanish Consolider-Ingenio 2010 Program CPAN (CSD2007-00042) and by Generalitat Valenciana Prometeo/2009/129.
%
%\bibliographystyle{unsrt}
%\bibliography{../../multiq6}

\begin{thebibliography}{10}%
\makeatletter
\providecommand \@ifxundefined [1]{%
 \ifx #1\undefined \expandafter \@firstoftwo
 \else \expandafter \@secondoftwo
\fi
}%
\providecommand \@ifnum [1]{%
 \ifnum #1\expandafter \@firstoftwo
 \else \expandafter \@secondoftwo
\fi
}%
\providecommand \enquote [1]{``#1''}%
\providecommand \bibnamefont  [1]{#1}%
\providecommand \bibfnamefont [1]{#1}%
\providecommand \citenamefont [1]{#1}%
\providecommand\href[0]{\@sanitize\@href}%
\providecommand\@href[1]{\endgroup\@@startlink{#1}\endgroup\@@href}%
\providecommand\@@href[1]{#1\@@endlink}%
\providecommand \@sanitize [0]{\begingroup\catcode`\&12\catcode`\#12\relax}%
\@ifxundefined \pdfoutput {\@firstoftwo}{%
 \@ifnum{\z@=\pdfoutput}{\@firstoftwo}{\@secondoftwo}%
}{%
 \providecommand\@@startlink[1]{\leavevmode}%
 \providecommand\@@endlink[0]{}%
}{%
 \providecommand\@@startlink[1]{%
  \leavevmode
  \pdfstartlink
   attr{/Border[0 0 1 ]/H/I/C[0 1 1]}%
   user{/Subtype/Link/A<</Type/Action/S/URI/URI(#1)>>}%
  \relax
 }%
 \providecommand\@@endlink[0]{\pdfendlink}%
}%
\providecommand \url  [0]{\begingroup\@sanitize \@url }%
\providecommand \@url [1]{\endgroup\@href {#1}{\urlprefix}}%
\providecommand \urlprefix [0]{URL }%
\providecommand \Eprint[0]{\href }%
\@ifxundefined \urlstyle {%
  \providecommand \doi [1]{doi:\discretionary{}{}{}#1}%
}{%
  \providecommand \doi [0]{doi:\discretionary{}{}{}\begingroup
  \urlstyle{rm}\Url }%
}%
\providecommand \doibase [0]{http://dx.doi.org/}%
\providecommand \Doi[1]{\href{\doibase#1}}%
\providecommand \bibAnnote [3]{%
  \BibitemShut{#1}%
  \begin{quotation}\noindent
    \textsc{Key:}\ #2\\\textsc{Annotation:}\ #3%
  \end{quotation}%
}%
\providecommand \bibAnnoteFile [2]{%
  \IfFileExists{#2}{\bibAnnote {#1} {#2} {\input{#2}}}{}%
}%
\providecommand \typeout [0]{\immediate \write \m@ne }%
\providecommand \selectlanguage [0]{\@gobble}%
\providecommand \bibinfo [0]{\@secondoftwo}%
\providecommand \bibfield [0]{\@secondoftwo}%
\providecommand \translation [1]{[#1]}%
\providecommand \BibitemOpen[0]{}%
\providecommand \bibitemStop [0]{}%
\providecommand \bibitemNoStop [0]{.\EOS\space}%
\providecommand \EOS [0]{\spacefactor3000\relax}%
\providecommand \BibitemShut [1]{\csname bibitem#1\endcsname}%
%</preamble>
\bibitem{Maiani:2004vq}%
  \BibitemOpen
  \bibfield{author}{%
  \bibinfo {author} {\bibfnamefont{L.}~\bibnamefont{Maiani}}, \bibinfo {author}
  {\bibfnamefont{F.}~\bibnamefont{Piccinini}}, \bibinfo {author}
  {\bibfnamefont{A.~D.}\ \bibnamefont{Polosa}},\ and\ \bibinfo {author}
  {\bibfnamefont{V.}~\bibnamefont{Riquer}},\ }%
  \bibfield{journal}{%
  \bibinfo {journal} {Phys. Rev.}\ }%
  \textbf{\bibinfo {volume} {D71}},\ \bibinfo {pages} {014028} (\bibinfo {year}
  {2005}),\ \Eprint{http://arxiv.org/abs/hep-ph/0412098}{hep-ph/0412098}%
  \bibAnnoteFile{NoStop}{Maiani:2004vq}%
%%CITATION = HEP-PH 0412098;%%
\bibitem{Ebert:2005nc}%
  \BibitemOpen
  \bibfield{author}{%
  \bibinfo {author} {\bibfnamefont{D.}~\bibnamefont{Ebert}}, \bibinfo {author}
  {\bibfnamefont{R.}~\bibnamefont{Faustov}},\ and\ \bibinfo {author}
  {\bibfnamefont{V.}~\bibnamefont{Galkin}},\ }%
  \bibfield{journal}{%
  \Doi{10.1016/j.physletb.2006.01.026}{\bibinfo {journal} {Phys.Lett.}}\ }%
  \textbf{\bibinfo {volume} {B634}},\ \bibinfo {pages} {214} (\bibinfo {year}
  {2006}),\ \Eprint{http://arxiv.org/abs/hep-ph/0512230}{arXiv:hep-ph/0512230
  [hep-ph]}%
  \bibAnnoteFile{NoStop}{Ebert:2005nc}%
%%CITATION = HEP-PH/0512230;%%
\bibitem{Dubnicka:2010kz}%
  \BibitemOpen
  \bibfield{author}{%
  \bibinfo {author} {\bibfnamefont{S.}~\bibnamefont{Dubnicka}}, \bibinfo
  {author} {\bibfnamefont{A.~Z.}\ \bibnamefont{Dubnickova}}, \bibinfo {author}
  {\bibfnamefont{M.~A.}\ \bibnamefont{Ivanov}},\ and\ \bibinfo {author}
  {\bibfnamefont{J.~G.}\ \bibnamefont{Korner}},\ }%
  \bibfield{journal}{%
  \Doi{10.1103/PhysRevD.81.114007}{\bibinfo {journal} {Phys.Rev.}}\ }%
  \textbf{\bibinfo {volume} {D81}},\ \bibinfo {pages} {114007} (\bibinfo {year}
  {2010}),\ \Eprint{http://arxiv.org/abs/1004.1291}{arXiv:1004.1291 [hep-ph]}%
  \bibAnnoteFile{NoStop}{Dubnicka:2010kz}%
%%CITATION = ARXIV:1004.1291;%%
\bibitem{Carlson:1991zt}%
  \BibitemOpen
  \bibfield{author}{%
  \bibinfo {author} {\bibfnamefont{J.}~\bibnamefont{Carlson}}\ and\ \bibinfo
  {author} {\bibfnamefont{V.~R.}\ \bibnamefont{Pandharipande}},\ }%
  \bibfield{journal}{%
  \bibinfo {journal} {Phys. Rev.}\ }%
  \textbf{\bibinfo {volume} {D43}},\ \bibinfo {pages} {1652} (\bibinfo {year}
  {1991})%
  \bibAnnoteFile{NoStop}{Carlson:1991zt}%
%%CITATION = PHRVA,D43,1652;%%
\bibitem{Vijande:2007ix}%
  \BibitemOpen
  \bibfield{author}{%
  \bibinfo {author} {\bibfnamefont{J.}~\bibnamefont{Vijande}}, \bibinfo
  {author} {\bibfnamefont{A.}~\bibnamefont{Valcarce}},\ and\ \bibinfo {author}
  {\bibfnamefont{J.~M.}\ \bibnamefont{Richard}},\ }%
  \bibfield{journal}{%
  \Doi{10.1103/PhysRevD.76.114013}{\bibinfo {journal} {Phys. Rev.}}\ }%
  \textbf{\bibinfo {volume} {D76}},\ \bibinfo {pages} {114013} (\bibinfo {year}
  {2007}),\ \Eprint{http://arxiv.org/abs/0707.3996}{arXiv:0707.3996 [hep-ph]}%
  \bibAnnoteFile{NoStop}{Vijande:2007ix}%
%%CITATION = 0707.3996;%%
\bibitem{Richard:2009rp}%
  \BibitemOpen
  \bibfield{author}{%
  \bibinfo {author} {\bibfnamefont{J.-M.}\ \bibnamefont{Richard}},\ }%
  \bibfield{journal}{%
  \Doi{10.1103/PhysRevC.81.015205}{\bibinfo {journal} {Phys. Rev.}}\ }%
  \textbf{\bibinfo {volume} {C81}},\ \bibinfo {pages} {015205} (\bibinfo {year}
  {2010}),\ \Eprint{http://arxiv.org/abs/0908.2944}{arXiv:0908.2944 [hep-ph]}%
  \bibAnnoteFile{NoStop}{Richard:2009rp}%
%%CITATION = 0908.2944;%%
\bibitem{Vijande:2011im}%
  \BibitemOpen
  \bibfield{author}{%
  \bibinfo {author} {\bibfnamefont{J.}~\bibnamefont{Vijande}}, \bibinfo
  {author} {\bibfnamefont{A.}~\bibnamefont{Valcarce}},\ and\ \bibinfo {author}
  {\bibfnamefont{J.-M.}\ \bibnamefont{Richard}},\ }%
  \bibfield{journal}{%
  \Doi{10.1103/PhysRevD.85.014019}{\bibinfo {journal} {Phys.Rev.}}\ }%
  \textbf{\bibinfo {volume} {D85}},\ \bibinfo {pages} {014019} (\bibinfo {year}
  {2012}),\ \Eprint{http://arxiv.org/abs/1111.5921}{arXiv:1111.5921 [hep-ph]}%
  \bibAnnoteFile{NoStop}{Vijande:2011im}%
%%CITATION = ARXIV:1111.5921;%%
\bibitem{Chan:1977st}%
  \BibitemOpen
  \bibfield{author}{%
  \bibinfo {author} {\bibfnamefont{H.-M.}\ \bibnamefont{Chan}}\ and\ \bibinfo
  {author} {\bibfnamefont{H.}~\bibnamefont{{H\o g\aa{}sen}}},\ }%
  \bibfield{journal}{%
  \bibinfo {journal} {Phys. Lett.}\ }%
  \textbf{\bibinfo {volume} {B72}},\ \bibinfo {pages} {121} (\bibinfo {year}
  {1977})%
  \bibAnnoteFile{NoStop}{Chan:1977st}%
%%CITATION = PHLTA,B72,121;%%
\bibitem{Chan:1978nk}%
  \BibitemOpen
  \bibfield{author}{%
  \bibinfo {author} {\bibfnamefont{H.-M.}\ \bibnamefont{Chan}} \emph{et~al.},\
  }%
  \bibfield{journal}{%
  \bibinfo {journal} {Phys. Lett.}\ }%
  \textbf{\bibinfo {volume} {B76}},\ \bibinfo {pages} {634} (\bibinfo {year}
  {1978})%
  \bibAnnoteFile{NoStop}{Chan:1978nk}%
%%CITATION = PHLTA,B76,634;%%
\bibitem{Vijande:2009ac}%
  \BibitemOpen
  \bibfield{author}{%
  \bibinfo {author} {\bibfnamefont{J.}~\bibnamefont{Vijande}}\ and\ \bibinfo
  {author} {\bibfnamefont{A.}~\bibnamefont{Valcarce}},\ }%
  \bibfield{journal}{%
  \Doi{10.3390/sym1020155}{\bibinfo {journal} {Symmetry}}\ }%
  \textbf{\bibinfo {volume} {1}},\ \bibinfo {pages} {155} (\bibinfo {year}
  {2009}),\ \Eprint{http://arxiv.org/abs/0912.3605}{arXiv:0912.3605 [hep-ph]}%
  \bibAnnoteFile{NoStop}{Vijande:2009ac}%
%%CITATION = ARXIV:0912.3605;%%
\bibitem{Lipkin:1986dw}%
  \BibitemOpen
  \bibfield{author}{%
  \bibinfo {author} {\bibfnamefont{H.~J.}\ \bibnamefont{Lipkin}},\ }%
  \bibfield{journal}{%
  \bibinfo {journal} {Phys. Lett.}\ }%
  \textbf{\bibinfo {volume} {B172}},\ \bibinfo {pages} {242} (\bibinfo {year}
  {1986})%
  \bibAnnoteFile{NoStop}{Lipkin:1986dw}%
%%CITATION = PHLTA,B172,242;%%
\bibitem{Brink:1994ic}%
  \BibitemOpen
  \bibfield{author}{%
  \bibinfo {author} {\bibfnamefont{D.}~\bibnamefont{Brink}}\ and\ \bibinfo
  {author} {\bibfnamefont{F.}~\bibnamefont{Stancu}},\ }%
  \bibfield{journal}{%
  \Doi{10.1103/PhysRevD.49.4665}{\bibinfo {journal} {Phys.Rev.}}\ }%
  \textbf{\bibinfo {volume} {D49}},\ \bibinfo {pages} {4665} (\bibinfo {year}
  {1994})%
  \bibAnnoteFile{NoStop}{Brink:1994ic}%
%%CITATION = PHRVA,D49,4665;%%
\bibitem{Bali:2000un}%
  \BibitemOpen
  \bibfield{author}{%
  \bibinfo {author} {\bibfnamefont{G.~S.}\ \bibnamefont{Bali}},\ }%
  \bibfield{journal}{%
  \Doi{10.1103/PhysRevD.62.114503}{\bibinfo {journal} {Phys.Rev.}}\ }%
  \textbf{\bibinfo {volume} {D62}},\ \bibinfo {pages} {114503} (\bibinfo {year}
  {2000}),\ \Eprint{http://arxiv.org/abs/hep-lat/0006022}{arXiv:hep-lat/0006022
  [hep-lat]}%
  \bibAnnoteFile{NoStop}{Bali:2000un}%
%%CITATION = HEP-LAT/0006022;%%
\bibitem{Korner:1994nh}%
  \BibitemOpen
  \bibfield{author}{%
  \bibinfo {author} {\bibfnamefont{J.}~\bibnamefont{K{\"o}rner}}, \bibinfo
  {author} {\bibfnamefont{M.}~\bibnamefont{Kramer}},\ and\ \bibinfo {author}
  {\bibfnamefont{D.}~\bibnamefont{Pirjol}},\ }%
  \bibfield{journal}{%
  \Doi{10.1016/0146-6410(94)90053-1}{\bibinfo {journal} {Prog. Part. Nucl.
  Phys.}}\ }%
  \textbf{\bibinfo {volume} {33}},\ \bibinfo {pages} {787} (\bibinfo {year}
  {1994}),\ \Eprint{http://arxiv.org/abs/hep-ph/9406359}{arXiv:hep-ph/9406359
  [hep-ph]}%
  \bibAnnoteFile{NoStop}{Korner:1994nh}%
%%CITATION = HEP-PH/9406359;%%
\bibitem{Fleck:1989mb}%
  \BibitemOpen
  \bibfield{author}{%
  \bibinfo {author} {\bibfnamefont{S.}~\bibnamefont{Fleck}}\ and\ \bibinfo
  {author} {\bibfnamefont{J.~M.}\ \bibnamefont{Richard}},\ }%
  \bibfield{journal}{%
  \bibinfo {journal} {Prog. Theor. Phys.}\ }%
  \textbf{\bibinfo {volume} {82}},\ \bibinfo {pages} {760} (\bibinfo {year}
  {1989})%
  \bibAnnoteFile{NoStop}{Fleck:1989mb}%
%%CITATION = PTPKA,82,760;%%
\bibitem{Quigg:1979vr}%
  \BibitemOpen
  \bibfield{author}{%
  \bibinfo {author} {\bibfnamefont{C.}~\bibnamefont{Quigg}}\ and\ \bibinfo
  {author} {\bibfnamefont{J.~L.}\ \bibnamefont{Rosner}},\ }%
  \bibfield{journal}{%
  \Doi{10.1016/0370-1573(79)90095-4}{\bibinfo {journal} {Phys.Rept.}}\ }%
  \textbf{\bibinfo {volume} {56}},\ \bibinfo {pages} {167} (\bibinfo {year}
  {1979})%
  \bibAnnoteFile{NoStop}{Quigg:1979vr}%
%%CITATION = PRPLC,56,167;%%
\bibitem{Hyodo:2012pm}%
  \BibitemOpen
  \bibfield{author}{%
  \bibinfo {author} {\bibfnamefont{T.}~\bibnamefont{Hyodo}}, \bibinfo {author}
  {\bibfnamefont{Y.-R.}\ \bibnamefont{Liu}}, \bibinfo {author}
  {\bibfnamefont{M.}~\bibnamefont{Oka}}, \bibinfo {author}
  {\bibfnamefont{K.}~\bibnamefont{Sudoh}},\ and\ \bibinfo {author}
  {\bibfnamefont{S.}~\bibnamefont{Yasui}}}%
   (\bibinfo {year} {2012}),\
  \Eprint{http://arxiv.org/abs/1209.6207}{arXiv:1209.6207 [hep-ph]}%
  \bibAnnoteFile{NoStop}{Hyodo:2012pm}%
%%CITATION = ARXIV:1209.6207;%%
\bibitem{Nielsen:2009uh}%
  \BibitemOpen
  \bibfield{author}{%
  \bibinfo {author} {\bibfnamefont{M.}~\bibnamefont{Nielsen}}, \bibinfo
  {author} {\bibfnamefont{F.~S.}\ \bibnamefont{Navarra}},\ and\ \bibinfo
  {author} {\bibfnamefont{S.~H.}\ \bibnamefont{Lee}},\ }%
  \bibfield{journal}{%
  \Doi{10.1016/j.physrep.2010.07.005}{\bibinfo {journal} {Phys.Rept.}}\ }%
  \textbf{\bibinfo {volume} {497}},\ \bibinfo {pages} {41} (\bibinfo {year}
  {2010}),\ \Eprint{http://arxiv.org/abs/0911.1958}{arXiv:0911.1958 [hep-ph]}%
  \bibAnnoteFile{NoStop}{Nielsen:2009uh}%
%%CITATION = ARXIV:0911.1958;%%
\bibitem{Valcarce:2012qwa}%
  \BibitemOpen
  \bibfield{author}{%
  \bibinfo {author} {\bibfnamefont{A.}~\bibnamefont{Valcarce}}, \bibinfo
  {author} {\bibfnamefont{T.}~\bibnamefont{Caram{\'e}s}},\ and\ \bibinfo
  {author} {\bibfnamefont{J.}~\bibnamefont{Vijande}},\ }%
  \bibfield{journal}{%
  \Doi{10.1007/s00601-012-0518-8}{\bibinfo {journal} {Few-Body Systems}},\
  \bibinfo {pages} {1}}%
   (\bibinfo {year} {2012}),\ ISSN \bibinfo {issn} {0177-7963},\
  \url{http://dx.doi.org/10.1007/s00601-012-0518-8}%
  \bibAnnoteFile{NoStop}{Valcarce:2012qwa}%
\bibitem{Lipkin:1998ew}%
  \BibitemOpen
  \bibfield{author}{%
  \bibinfo {author} {\bibfnamefont{H.}~\bibnamefont{Lipkin}},\ }%
  \bibfield{journal}{%
  \Doi{10.1016/S0370-2693(98)00707-2}{\bibinfo {journal} {Phys.Lett.}}\ }%
  \textbf{\bibinfo {volume} {B433}},\ \bibinfo {pages} {117} (\bibinfo {year}
  {1998})%
  \bibAnnoteFile{NoStop}{Lipkin:1998ew}%
%%CITATION = PHLTA,B433,117;%%
\end{thebibliography}
%\end{document}
%
%Merlin.mbs v4.21 2009-07-09.
%

\end{document}